\documentclass[runningheads]{llncs}

\usepackage[rightcaption]{sidecap}
\usepackage[bottom]{footmisc}
\usepackage{graphicx}
\graphicspath{ {./images/} }
\usepackage{float} 
\usepackage{booktabs}
\usepackage{tikz}

\DeclareRobustCommand\circled[1]{\tikz[baseline=(char.base)]{
            \node[shape=circle,draw,inner sep=1pt] (char) {#1};}}

\begin{document}
\title{Discovering care pathways for multi-morbid patients using event graphs}
%
%
\author{Milad Naeimaei Aali\inst{1} \and
Felix Mannhardt\inst{2}\and
Pieter Jelle Toussaint\inst{1}}
\authorrunning{M. Naeimaei Aali}
%
\institute {
Norwegian University of Science and Technology, Trondheim, Norway 
\and Eindhoven University of Technology, Eindhoven, Netherlands
}
\maketitle              
\begin{abstract} Patients suffering from multiple diseases (multi-morbid patients) often have complex clinical pathways. They are diagnosed and treated by different specialties and undergo other clinical actions related to various diagnoses. Coordination of care for these patients is often challenging, and it would be of great benefit to get better insight into how the clinical pathways develop in reality. Discovering these pathways using traditional process mining techniques and standard event logs may be difficult because the patient is involved in several highly independent clinical processes. Our objective is to explore the potential of analyzing these pathways using an event log representation reflecting the independent clinical processes. Our main research question is: How can we identify valuable insights by using a multi-entity event data representation for clinical pathways of multi-morbid patients? Our method was built on the idea to represent multiple entities in event logs as event graphs. The MIMIC-III dataset was used to evaluate the feasibility of this approach. Several clinical entities were identified and then mapped into an event graph. Finally, multi-entity directly follows graphs were discovered by querying the event graph visualizing them. Our result shows that paths involving multiple entities include traditional process mining concepts not for one clinical process but all involved processes. In addition, the relationship between activities of different clinical processes, which was not recognizable in traditional models, is visible in the event graph representation. 


\keywords{Health care \and Multi morbidity \and Multi-entity Process Mining}
\end{abstract}
\section{Introduction}

Based on the UN annual report, the number of older people is envisaged to be nearly 2.1 billion by 2050, growing to a size more than twice as large as in 2017~\cite{UN}. As a result of the aging population, it is expected that "multi-morbidity" is going to increase~\cite{Marengoni}. Multi-morbidity refers to any co-occurrence of conditions in the same person~\cite{Marengoni}. Sometimes the term "co-morbidity" is used instead of multi-morbidity, while the term co-morbidity is defined as the combination of extra disorders besides an index disease~\cite{Marengoni}. The treatment of multi-morbid patients is a complicated task since they generate several challenges. These include recognizing signs and symptoms of different illnesses, managing multiple medications and treatments, interacting between various health conditions, and allocating resources by medical centers. These lead us to develop care pathways for patients with multi-morbidity in a way that overcomes these challenges.

Care pathways, as one of the central tools used in healthcare, can be described as a straightforward statement of the aims, a representation of the interactions between the health's resources and patients, or a description of roles, sequential decisions, and activities related to the care process~\cite{Schrijvers}. The primary goal of care pathways is reducing variability in the treatment of diseases~\cite{Fernandez}. Since care pathways are a set of time-framed events focusing on a specific situation that delivers guidance about how to deal with conditions that appear during treatment's processes~\cite{Fernandez}. It can be itself considered as a process which is a sequence of events with a common goal~\cite{Wil}.

Processes can be graphically represented by process models~\cite{Wil} which explain responsibilities, inspect compliance, predict performance using simulation~\cite{Alessandro}, manage complexity, reduce variation, and enhance coordination~\cite{Wil} in processes. Discovering process models or process discovery from event logs is one of the main tasks in process mining. Event logs contain sequences of events recorded from information systems. Any registered event refers to at least (1) an activity (i.e., a well-defined step in the process), (2) a case or process instance representing a single entity, (3) a unique timestamp. Logs fulfilling these requirements are called single-dimensional or single-entity event data~\cite{Fahland}, which an example of this type of log was shown in Table~\ref{tab1}. Single entity event data also can refer to properties (e.g., the person executing or initiating the activity)~\cite{Wil}. 

If we want to satisfy all practitioners in the healthcare sector and achieve a holistic process view for care pathways~\cite{Niels}, we should consider more than one clinical process of patients' care pathways. But, the standard type of event data forces us to deploy an event log for each clinical process of patient care pathways. On the other hand, if we have multi-entity event data, meaning events refer to multiple entities (e.g., each clinical process of a care pathway), relational databases and traditional process mining techniques are ineffective.

\begin{table}[tb]
\centering
\caption{Example of event log with single-dimensional or single-entity event data.}\label{tab1}
{\scriptsize
\begin{tabular}{lllll}
\toprule
Case Identifier &  Event & Timestamps & Property-X & Property-Y \\
\midrule
1 & a & 2013-10-29T05:00:00 & X1 & Y1\\
1 & b & 2013-10-30T06:00:00 & X2 & Y2\\
1 & c & 2013-10-31T07:00:00 & X3 & Y3\\
2 & a & 2013-10-01T08:00:00 & X4 & Y4\\
2 & c & 2013-10-19T09:00:00 & X5 & Y5\\
3 & a & 2013-10-29T06:00:00 & X6 & Y6\\
\bottomrule
\end{tabular}
}
\end{table}

This study explores the potential of analyzing care pathways for patients with multi-morbidity using a multi-entity event data representation reflecting the independent clinical processes. Our main research question is: \textit{How can we identify valuable insights by using a multi-entity event data representation for care pathways of multi-morbid patients?} The remainder of this research is structured as follows. Section 2 reviews state-of-the-art research about the use of multi-entity event data in process mining and how to represent and store them. Section 3 introduces MIMIC-III that is used to illustrate and validate our approach. In Section 4, we show how to build multi-entity event data for multi-morbid patients. In Section 5, we show preliminary results that are, then, discussed in Section 6. We conclude with an outlook on future work in Section 7.

\section{Related Work}

Multi-entity event data can not be stored in the same way as single-entity event data; furthermore, in this setting process discovery is not possible with traditional methods. In this section, we explore the related literature from several perspectives to select a good format for multi-morbid care pathways event data.

\subsection{Multi-entity event data}

In the approach of~\cite{Aalst}, known as object-centric process mining, each case notion is referred to as one object type (e.g., application and vacancy can be two case notions or two object types, and each of them has its own case identifiers). In that approach, events can refer to multiple case notions instead of referring to a single case notion. A process model is first discovered for all objects sequentially. Then, each directly-follows relation is labeled to its related object type. For example, if event-1 that is related to object-1 happened right before event-2 that is related to object-1, event-2 directly follows event-1, and so on.

Another type of multi-entity event data was proposed by~\cite{Fahland}. Based on~\cite{Fahland}, there does not need to be a single case notion, but events are related to one or more \emph{entities} of different entity types. Entities themselves can also be related to each other. The required input events have been shown in~\cite{Fahland} is similar to the one shown in Table~\ref{tab2}. Information about the relations may also be extracted from other sources, e.g., relational database keys. Process models can be discovered in a flexible manner per entity or for various combinations of several entities. Our event log format for storing multi-entity event data is based on this model.

\begin{table}[tb]
\centering
\caption{Excerpt of a event log with multi-entity event data relating to multiple entities that can be converted into an event graph representation~\cite{Fahland}.}\label{tab2}
{\scriptsize
\begin{tabular}{lllllll}
\toprule
Event & Timestamps & EntityTypeA & EntityTypeB & EntityTypeC & PropertyX & PropertyY \\
\midrule
a & 2013-10-29T05:00:00 & 1 & Origin 1 & Origin 4 & X1 & Y1\\
b & 2013-10-30T07:00:00 & 1 &  & Origin 4 & X4 & Y4\\
c & 2013-10-31T07:00:00 & 1 &  & Origin 5 & X5 & Y5\\
f & 2013-10-31T09:00:00 & 1 & Origin 1 & Origin 4 & X7 & Y7\\
a & 2013-10-01T08:00:00 & 2 & Origin 2 & Origin 4 & X2 & Y2\\
b & 2013-10-30T06:00:00 & 2 &  & Origin 4 & X3 & Y3\\
c & 2013-10-31T07:00:00 & 2 &  & Origin 5 & X5 & Y5\\
f & 2013-10-31T09:00:00 & 2 & Origin 2 & Origin 4 & X7 & Y7\\
a & 2013-10-29T05:00:00 & 3 & Origin 1 & Origin 4 & X1 & Y1\\
b & 2013-10-30T06:00:00 & 3 &  & Origin 4 & X3 & Y3\\
c & 2013-10-19T09:00:00 & 3 &  & Origin 5 & X6 & Y6\\
\bottomrule
\end{tabular}
}
\end{table}

\subsection{Storing Event data}

A classical approach for storing event data is using relational databases (RDBs). A relatively new approach is using an event graph which is a mathematical graph data structure that is built by converting relational database concepts to vertices, edges, nodes, and relationships~\cite{Fahland}. This leads to a natural representation of multi-entity event data and the possibility to discover multi-entity models by querying from event graphs.

A series of experiments were conducted in~\cite{Joishi} to compare the performance and efficacy of relational databases and event graphs, sho1higher capabilities of event graphs. Extracting multi-entity event data needs to flatten event data because only a single case notion can be chosen~\cite{Fahland} leading to traditional process mining. Additionally, a graph database can store all of the case notions of a multi-entity directly follows graph in only one graph~\cite{Fahland}. 

Recently, event graphs were deployed for storing data. The work in~\cite{Jalali} introduces an approach to store and retrieve single-entity event logs into/from graph databases. That approach defines how log files shall be stored in a graph database, and it also illustrates how directly follows graphs (DFG) can be calculated in the graph database. In another recent literature, task executions and routines in processes were classified and detected using event graphs~\cite{Klijn}. In that research, at first, the event log was transformed into an event graph. Then graph theory was used to detect task execution patterns and their changes over time. 

Converting multi-entity event data to an event graph was formalized in~\cite{Fahland} by conceptualizing event log, events, entities, and classes. Based on~\cite{Fahland} each event log has several events, and each event in one hand correlates to entities, and on the other hand, can be observed by classes. Meanwhile, the events can be related to each other if they directly follow each other. Entities can be related to each other based on the occurrence of their events. As well, the classes can follow each other by directly following relationships. Based on these reasons, in sum, an event graph seems to be a better approach compared to the relational database for storing multi-entity event data.

Vogelsang et al.~\cite{Vogelgesang} looks at process mining from multiple dimensions. Still, these dimensions are related to properties of cases such as region, age of patients, and not event data. In the approach, several single-entity event data, separated based on the difference between regions, ages, and so forth, were used. 

Overall, we found that the subject of using event graphs in a healthcare setting and, in particular, discovering care pathways from multi-entity event data using event graphs was not yet explored in previous literature.

\section{Multi-entity Event Data in MIMIC-III}

For evaluation of the feasibility of using event graphs for clinical pathways of multi-morbid patients, the MIMIC-III~\cite{MIMIC} is used. MIMIC-III is a freely accessible tertiary care database that involves information relating to patients admitted to critical care units (CCU) of Beth Israel Deaconess Medical Center in Boston, Massachusetts, during 2001 and 2012. Data from MIMIC-III were downloaded from several sources such as critical care monitoring information systems, bedside monitors, hospital and laboratory electronic health record databases, and social security administration. 

The ninth revision of the international statistical classification of diseases and related health problems (ICD-9) is widely used diagnostic coding system. Each ICD-9 code corresponds to a single diagnostic disease except the codes starting with E and V, which are related to external causes of injury and additional classification. We use the ICD-9 code system for specifying multi-morbid patients by considering patients with several ICD-9 codes as patients with multi-morbidity.

We use a subset of data from MIMIC-III. To extract event data from MIMIC-III, first, from \texttt{DIAGNOSES\_ICD} Table, values of \textit{icd9\_code} column, excluding codes start with E and V, were grouped by each distinct patient's hospital admission identifier (\textit{hadm\_id}). The \texttt{DIAGNOSES\_ICD} table involves patients identifiers (\textit{subject\_id}), patients hospital admission identifiers (\textit{hadm\_id}), the sequence order in which the ICD-9 diagnoses were made (\textit{seq\_id}), and ICD-9 (\textit{icd9\_code}). After that, the patient admission identifier was grouped by an collection of ICD codes as shown in Table~\ref{tab3}. Each row of Table~\ref{tab3}, shows the number of observances of a disease (or group of diseases), which has been coded by ICD-9 format, at the time of admission of patients to the hospital. If the first row of the table shows more than one disease, we consider them as multi-morbidity cases. Meanwhile, a patient can have several admission identifiers that show the patients admitted to the hospital several times at different times. 

From this initial look at a subset of the MIMIC-II dataset on multi-morbid patients, multiple entities can be identified, e.g., admissions, diseases (ICD codes), and so on. We now describe the relevant entities in detail and extract them to build an event graph representation.


\begin{table}[tb]
\centering
\caption{List of patient ICD code and its repetitive in patients.}\label{tab3}
{\scriptsize
\begin{tabular}{llllll}
\toprule
Diagnoses based on ICD codes &  Patients frequency & List of Patients (Admission IDs)\\
\midrule
7746 & 232 & A1 - A2 - ... - A232\\
7661 & 163 & B1 - B2 - ... - B163\\
7706 & 142 & C1 - C2 - ... - C142\\
76519 - 76528 & 99 & D1 - D2 - ... - D99\\
76518 - 76528 & 68 & E1 - E2 - ... - E68\\
77089 & 63 & F1 - F2 - ... - F63\\
... & ... & ...\\
\bottomrule
\end{tabular}
}
\end{table}

\section{Event Graphs for Multi-Morbid Patients Pathways}

This study explores how to analyze multi-entity event data for patients with multi-morbidity based on an event graph.  Based on our research question, a hypothesis for this research was formulated as follows: \textit{Applying event graph produces valuable insights when using multi-entity event data for clinical pathways of multi-morbid patients}. Our strategy is to design an experiment for the research to investigate this. This section describes the method we followed to investigate this question and build event graphs to discover care pathways for multi-morbid patients.

\subsection{Identifying and Extracting Entities}
Each distinct clinical process related to patients with multi-morbidity is called an entity. Since several clinical processes are involved in treating multi-morbid patients, entities can easily be identified by considering those clinical processes. We identified the following entities in the subset of the MIMIC-III dataset:

\begin{enumerate}
\item \textbf{Logistic}. This entity events contains admission, discharging, registering to Emergency department (ED), discharging from ED, In-hospital death (if died), calling-out request (when patients ready to discharge), and transferring between different services, care unit and wards. Six MIMIC-III tables were used to download this entity's events: \texttt{PATIENTS}, \texttt{ADMISSIONS}, \texttt{CALLOUT}, \texttt{SERVICES}, \texttt{ICUSTAYS}, \texttt{TRANSFERS}.

\item \textbf{Laboratory\_Measurement}. This entity contains events of the type abnormal laboratory measurements, Which play an essential role in diagnosing and treating patients' diseases. For extracting these events \textit{label}, \textit{value}, \textit{valueuom}, and \textit{flag} columns of \texttt{D\_LABITEMS}, and \texttt{LABEVENTS} tables were used. 

\item \textbf{Prescriptions}. This entity contains starting and ending timestamps of medication-related order entries, i.e., prescriptions such as the drug which is prescribed to the patient, its dose's value, form, and unit of medication, for extracting of this entity \texttt{PRESCRIPTIONS} table was used.

\item \textbf{Diagnosis}. This entity was related to the first event at the beginning of each time of patients admissions. It involves a group of ICD codes showing patients' diseases in each admission. \texttt{DIAGNOSES\_ICD} table relationship with other tables was used for downloading ICD codes of this entity.

\item \textbf{Admission}. In the end, the hospital admission identifier was appended to multi-entity event data. If an event is related to the NULL admission number, it is associated with the outpatient clinic.

\end{enumerate}

\noindent Table~\ref{tab4} shows an example of created multi-entity event data for patients identified 4900. It is possible to extract multi-entity event data for each row of Table~\ref{tab4}, while we consider the admission identifier or its equivalent patient identifier as a case identifier.

\begin{table}[tb]
\centering
\caption{Excerpt of an event log extracted from MIMIC-III with multiple entities. We abbreviate event labels in the remainder as follows: L\_Taken = Laboratory Test Taken, LAM = Laboratory Abnormal Measurement, CA = Coronary Atherosclerosis, DM = Diabetes Mellitus, HL = Hypercholesterolemia, HT = Hypertension, TBS = Transfers Between Services, TIW 27 = Transfer Into Ward: 27, HA = Hospital Admission.}\label{tab4}
{\scriptsize
\begin{tabular}{lllll}
\toprule
Patient Identifier &  Event & Timestamps & EntityType & Admission   \\
\midrule
Patient\_4900 & L\_Taken & 2013-10-29T05:00:00 & Lab. Measurement & Outpatient \\
Patient\_4900 & LAM & 2013-10-30T06:00:00 & Lab. Measurement & Outpatient \\
Patient\_4900 & CA DM HL HT & 2013-10-31T07:00:00 & Diagnosis & 115281\\
Patient\_4900 & TBS & 2013-10-01T08:00:00 & Logistic & 115281 \\
Patient\_4900 & TIW 27 & 2013-10-19T09:00:00 & Logistic & 115281\\
Patient\_4900 & HA & 2013-10-29T06:00:00 & Logistic & 174010\\
... & ... & ... & ... & ...\\
\bottomrule
\end{tabular}
}
\end{table}

\subsection{Building the Event Graph}

\begin{figure}[p]
    \centering
        \includegraphics[scale=0.24]{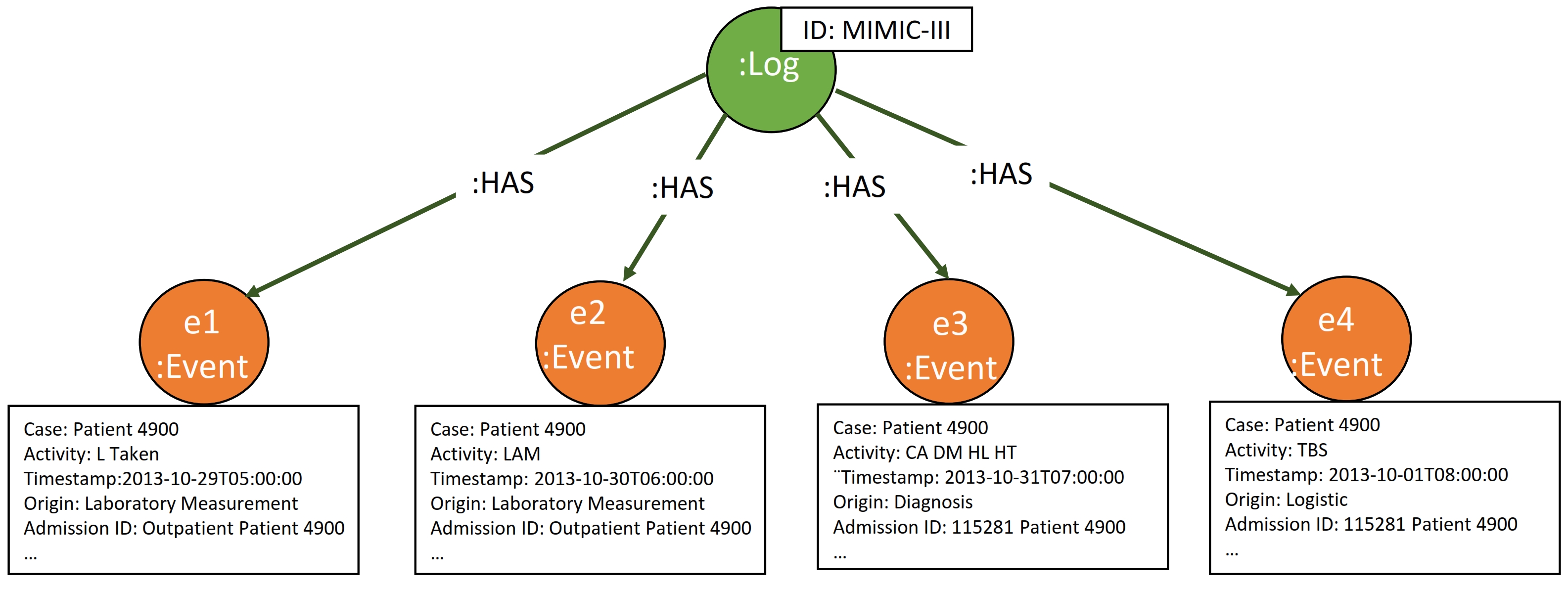}\newline
        \begin{minipage}[t]{.5\textwidth}%
            \includegraphics[scale=0.24]{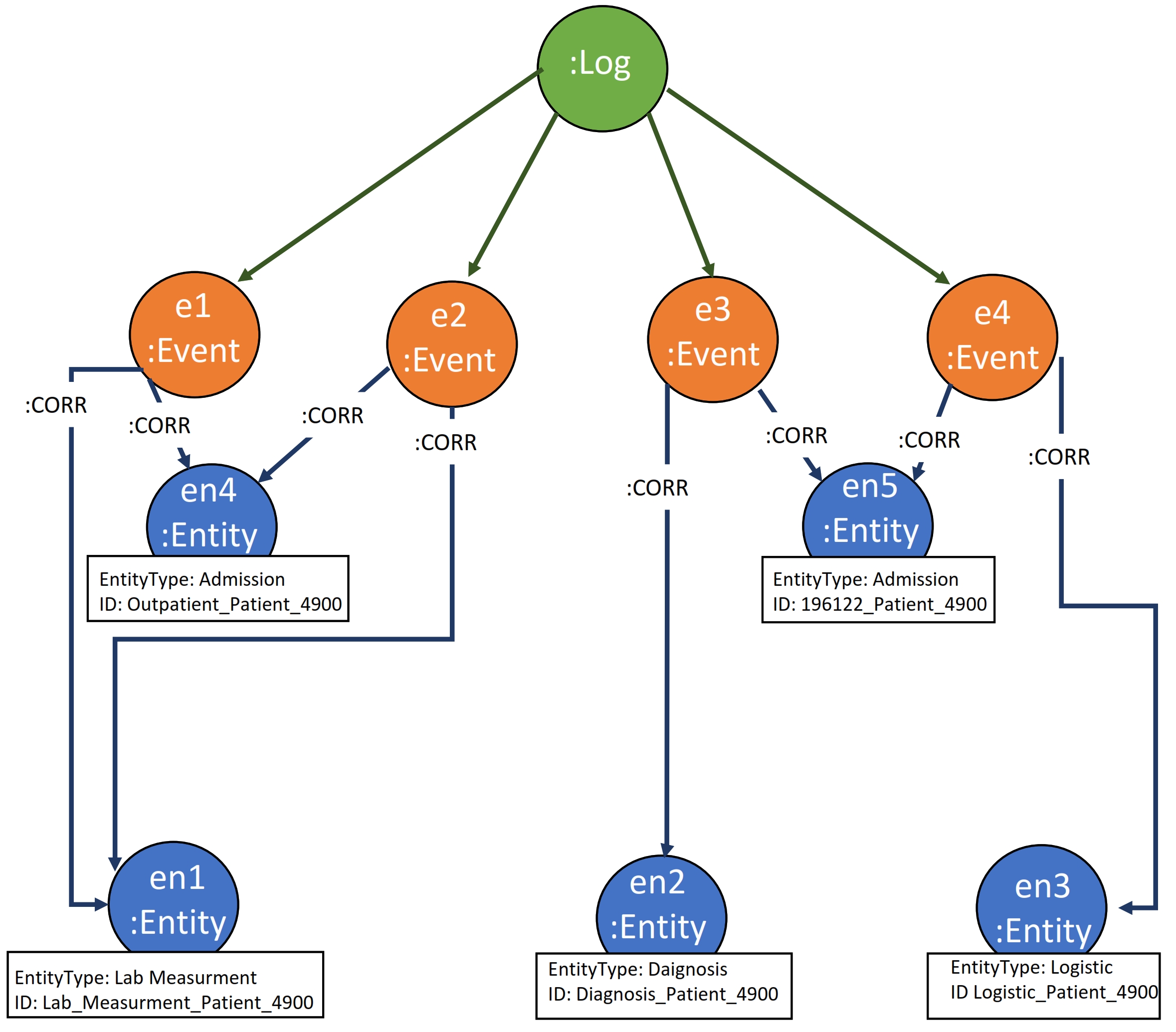} %
        \end{minipage}%
        \begin{minipage}[t]{.5\textwidth}%
        \includegraphics[scale=0.24]{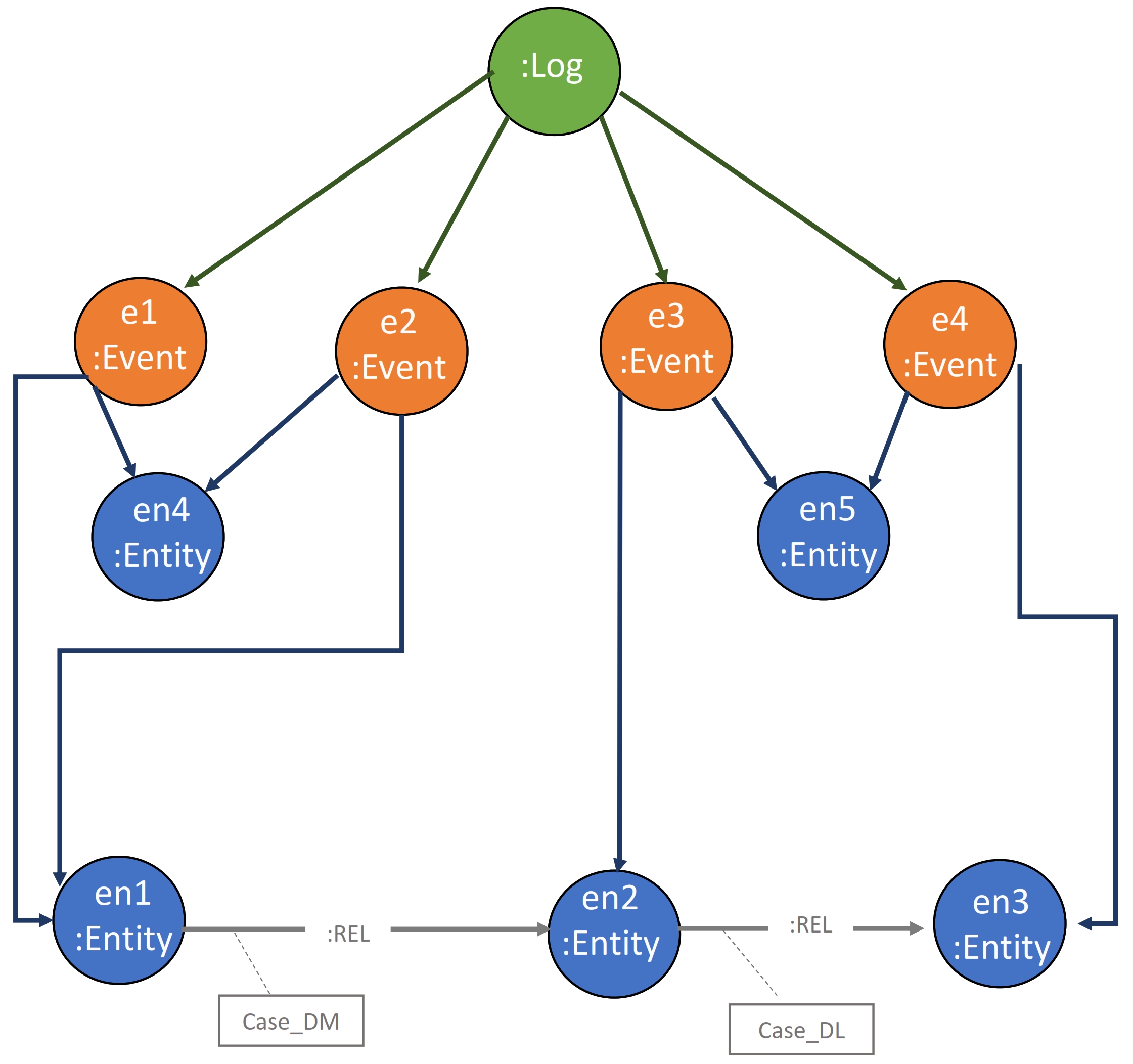}%
        \end{minipage}

    \caption{Graph creation for Patient\_4900: Steps \circled{1} (top), \circled{2} (bottom left), and \circled{3} (bottom right)}
    \label{fig:step123}  
\end{figure}

\begin{figure}[p]
    \centering
        \begin{minipage}[t]{.5\textwidth}%
            \includegraphics[scale=0.24]{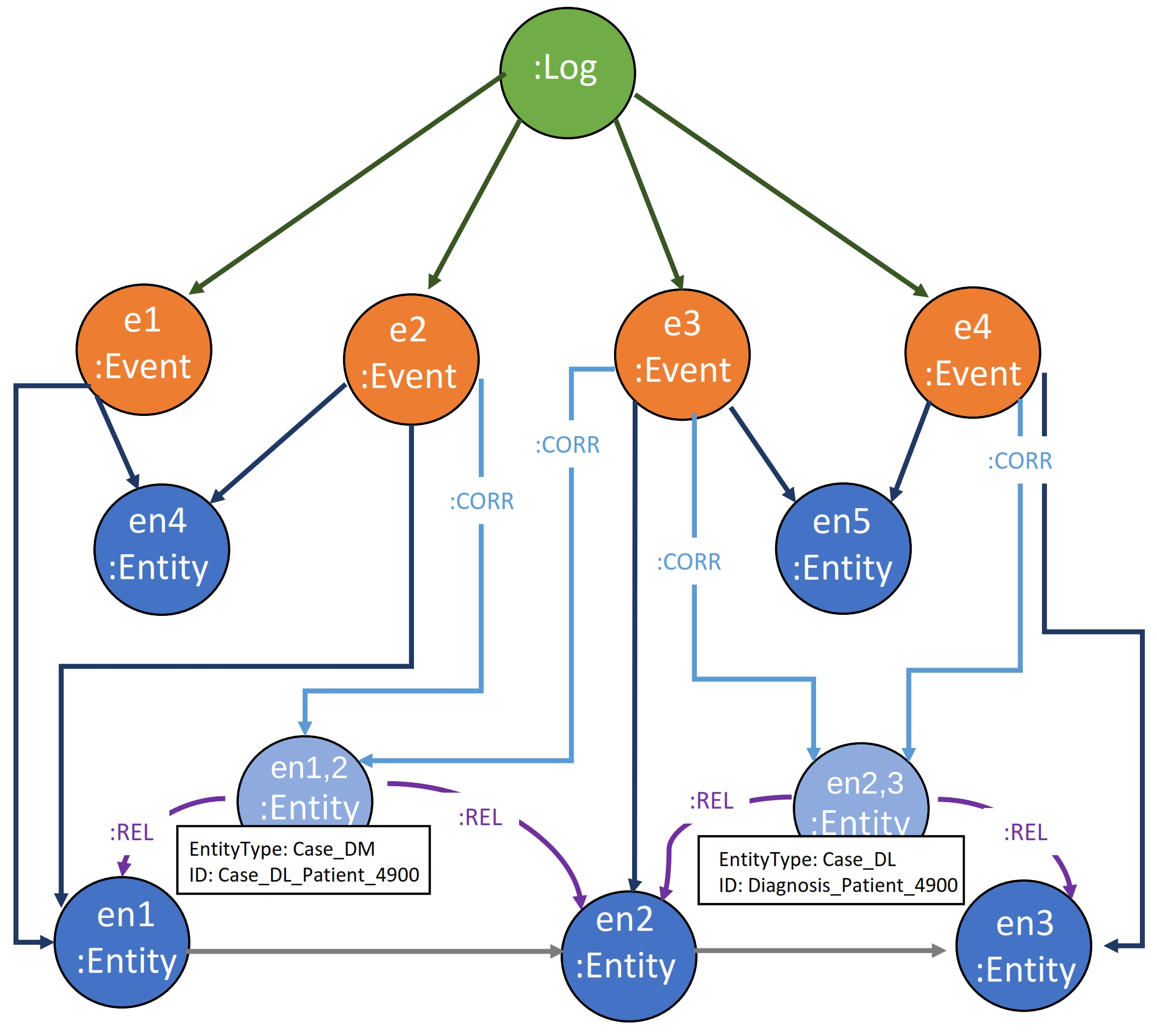} %
        \end{minipage}%
        \begin{minipage}[t]{.5\textwidth}%
            \includegraphics[scale=0.24]{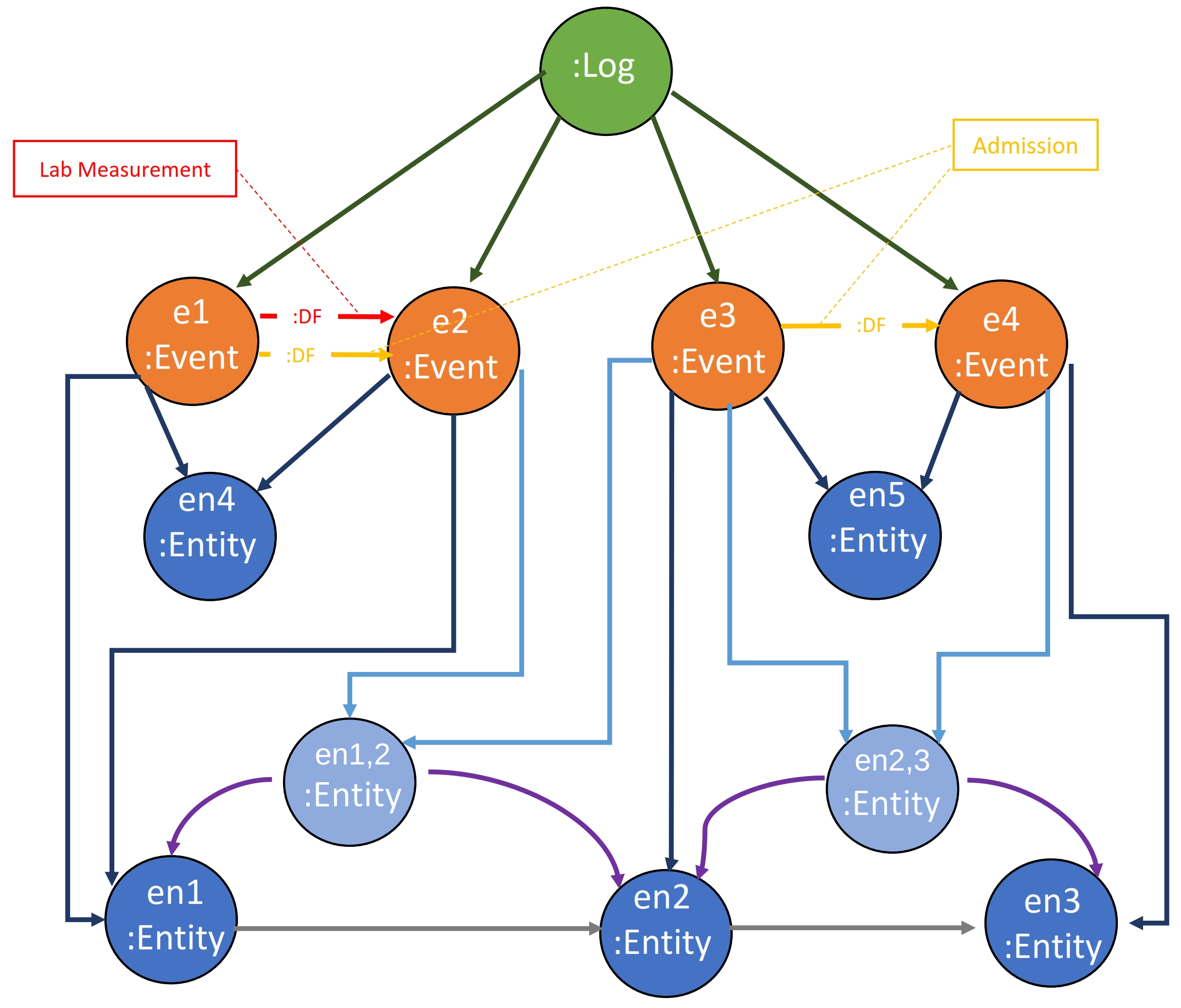}%
        \end{minipage}    

    \includegraphics[scale=0.24]{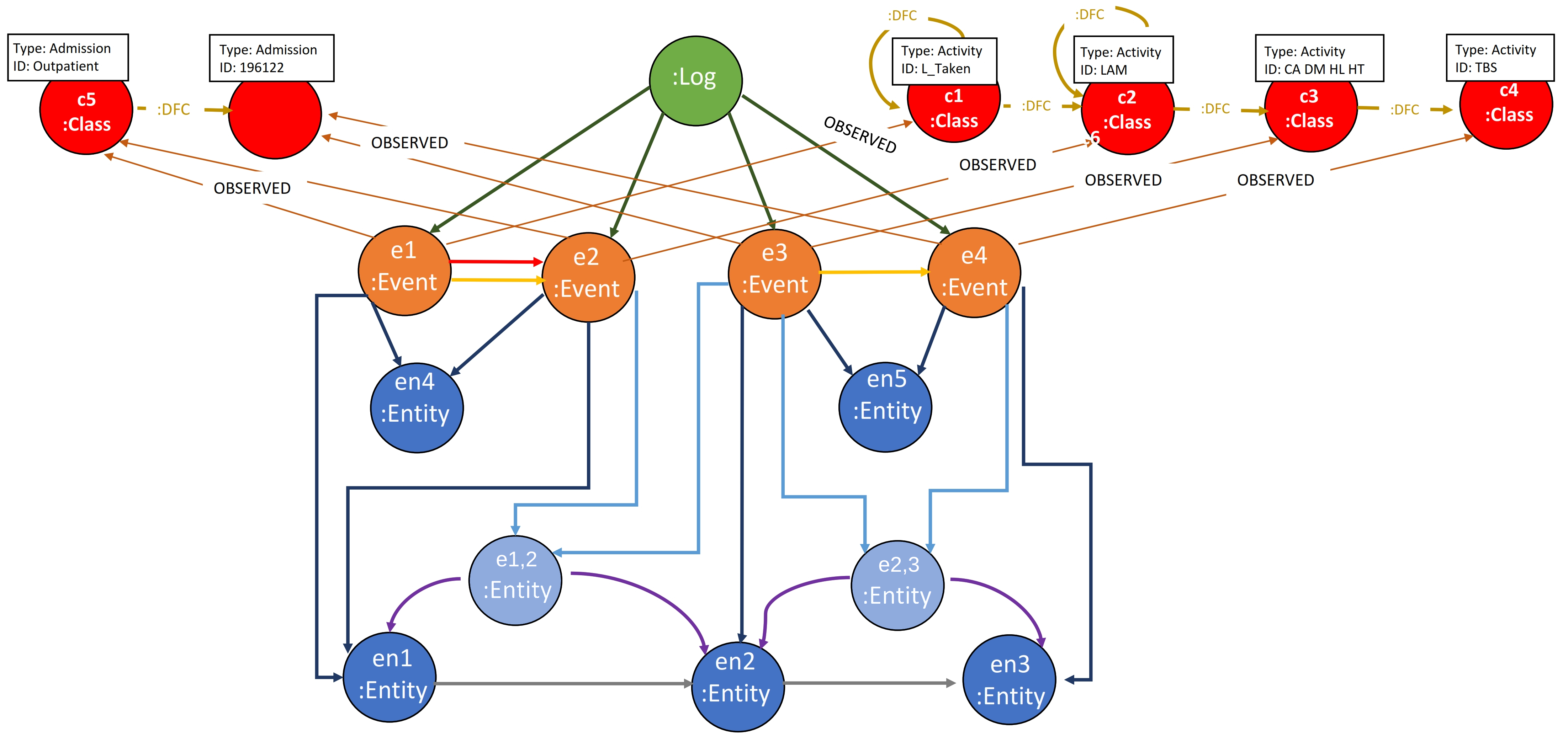}\newline

    \caption{Graph creation for Patient\_4900: Steps \circled{4} (top left), \circled{5} (top right), \circled{6} (bottom)}
    \label{fig:step456}
\end{figure}

We showed the steps we followed to create the event graph from the multi-entity event data based on the approach introduced in~\cite{Fahland} in Figures. \ref{fig:step123}, \ref{fig:step456}: \circled{1} Each record of the event log was converted to a node, called event node; then another node was created for the event log. After that, relationships from each event node to the log node was created. \circled{2} Nodes for the cases' entities and their properties, called entity nodes was generated, then each event node was correlated to its relative entity node. \circled{3} The entities nodes were related to each other based on their event's sequential occurrence. \circled{4} The relationship between the entities nodes were reified. \circled{5} Directly follows relation between the events node was created based on entities and properties, and \circled{6} Event class nodes and property class nodes were created respectively for distinct events and properties, and finally aggregated directly follows relationships for the event and property class nodes were created.

\section{Results of Application to MIMIC-III}

A preliminary evaluation of our approach relies on a qualitative discussion. We analyze the generated multi-entity directly follows graphs from the MIMIC-III database and evaluate to which extend they support our hypothesis. 
We implemented the event graph creation using Python and the Neo4J library and adapted the code provided by~\cite{Fahland} for our case\footnote{Available on \url{https://github.com/mnaeimaei/MIMICIII-Event-Graph}}. Multi-entity directly follows graphs were discovered by querying the event graph with CQL and visualized it with Graphviz.

\begin{figure}[!ht]
    \centering
    \includegraphics[scale=0.026]{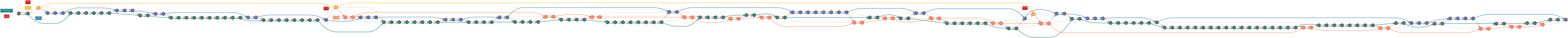}
    \includegraphics[scale=0.42]{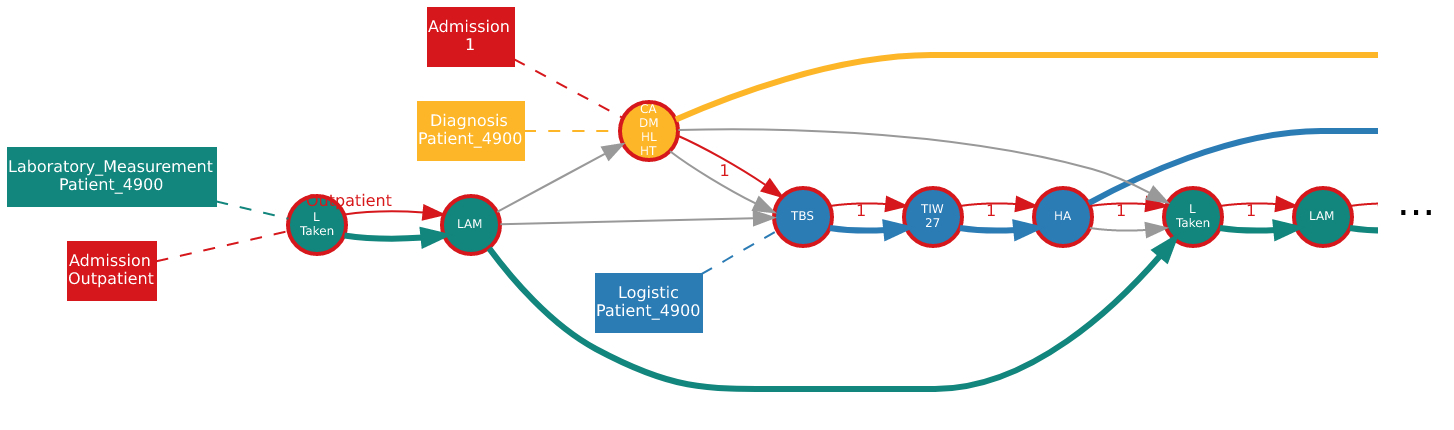}\newline
    \includegraphics[scale=0.42]{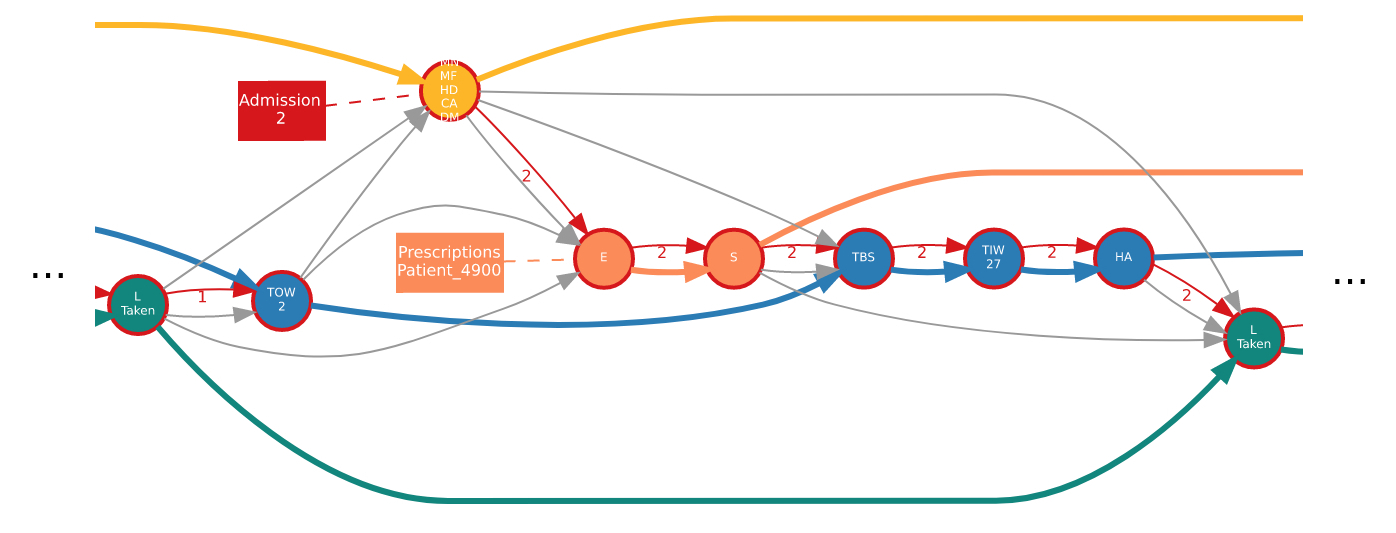}\newline
    \includegraphics[scale=0.42]{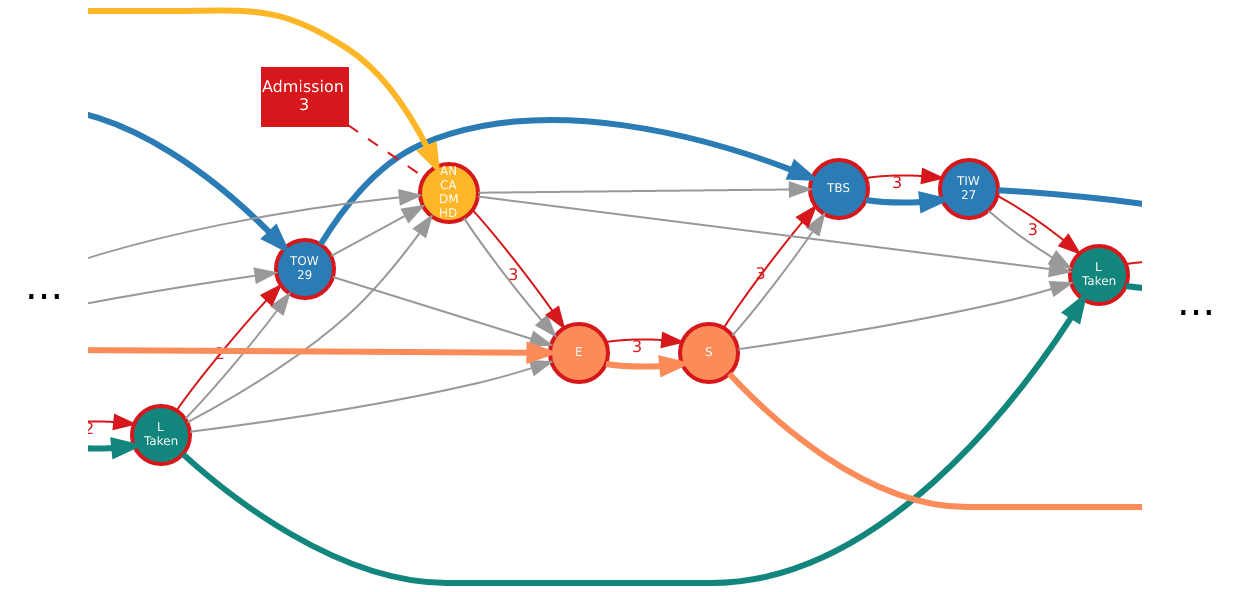}\newline
    \includegraphics[scale=0.42]{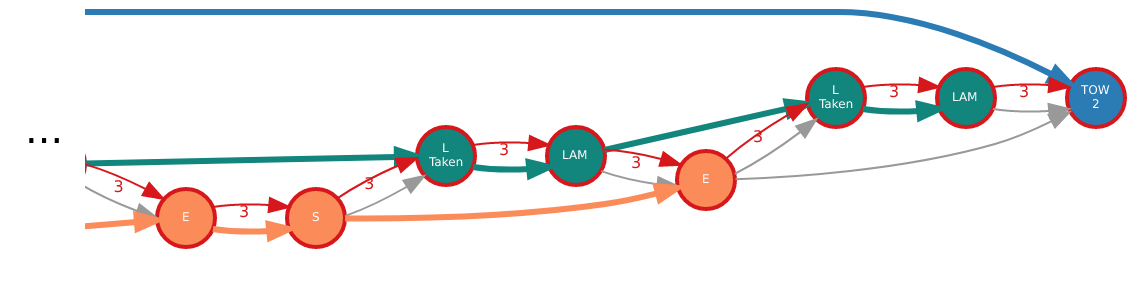}
    \caption{Multi-entity directly-follows graph for Patient\_4900.}
    \label{fig:one}
\end{figure}

\begin{figure}[!ht]
    \centering
    \includegraphics[scale=0.056]{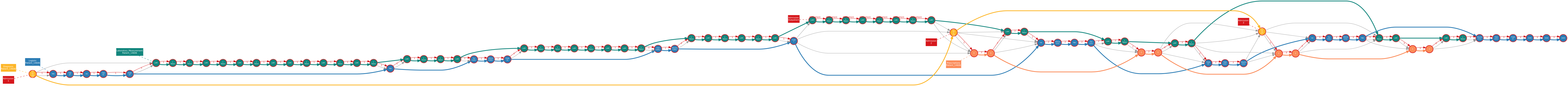}
    \includegraphics[scale=0.42]{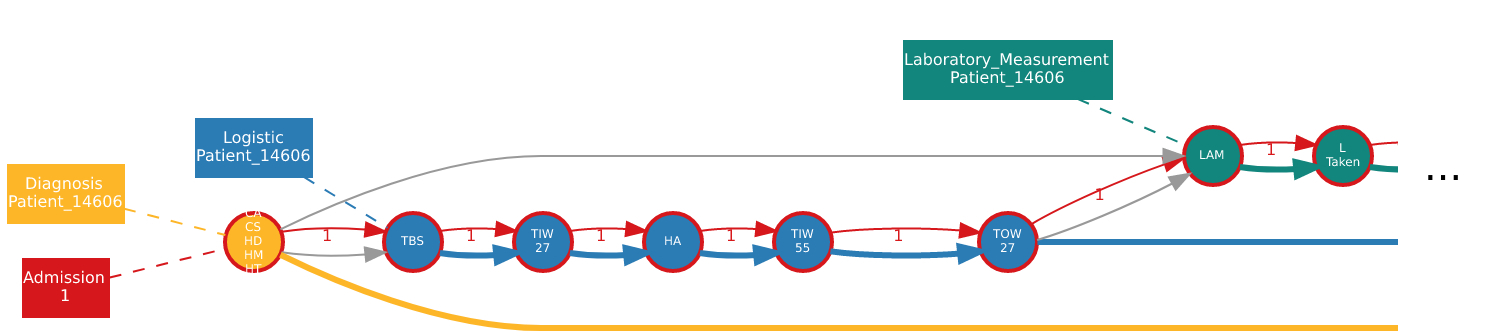}\newline
    \includegraphics[scale=0.42]{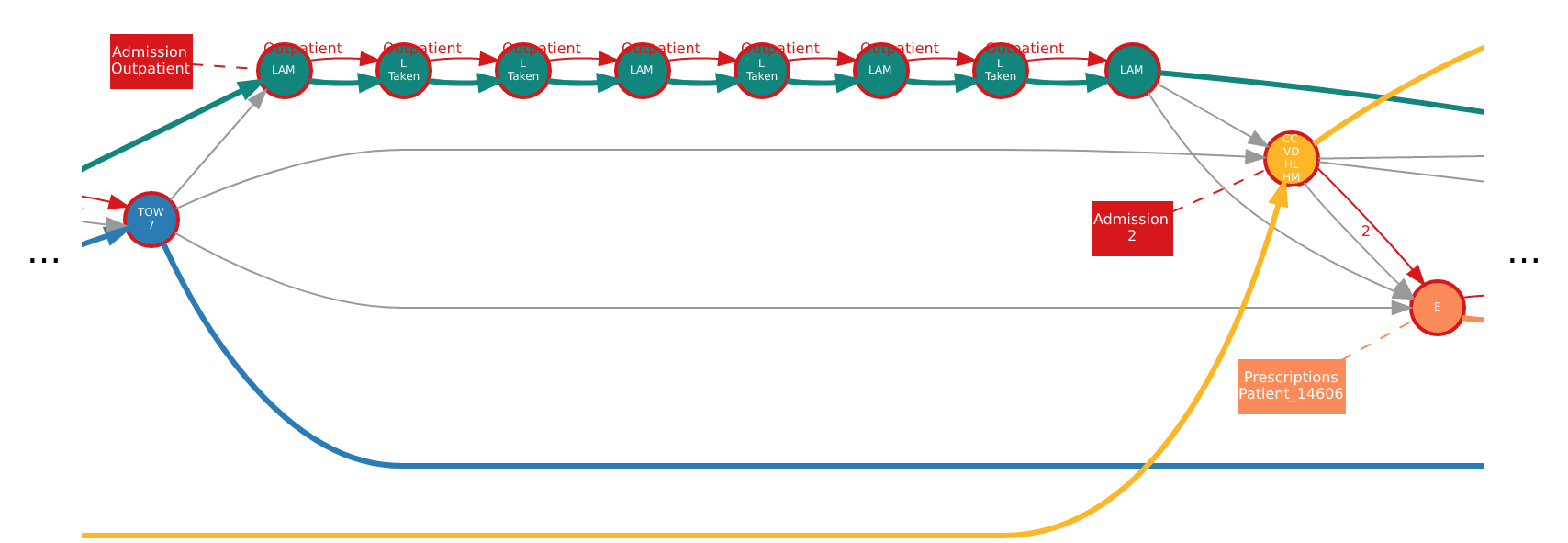}\newline
    \includegraphics[scale=0.42]{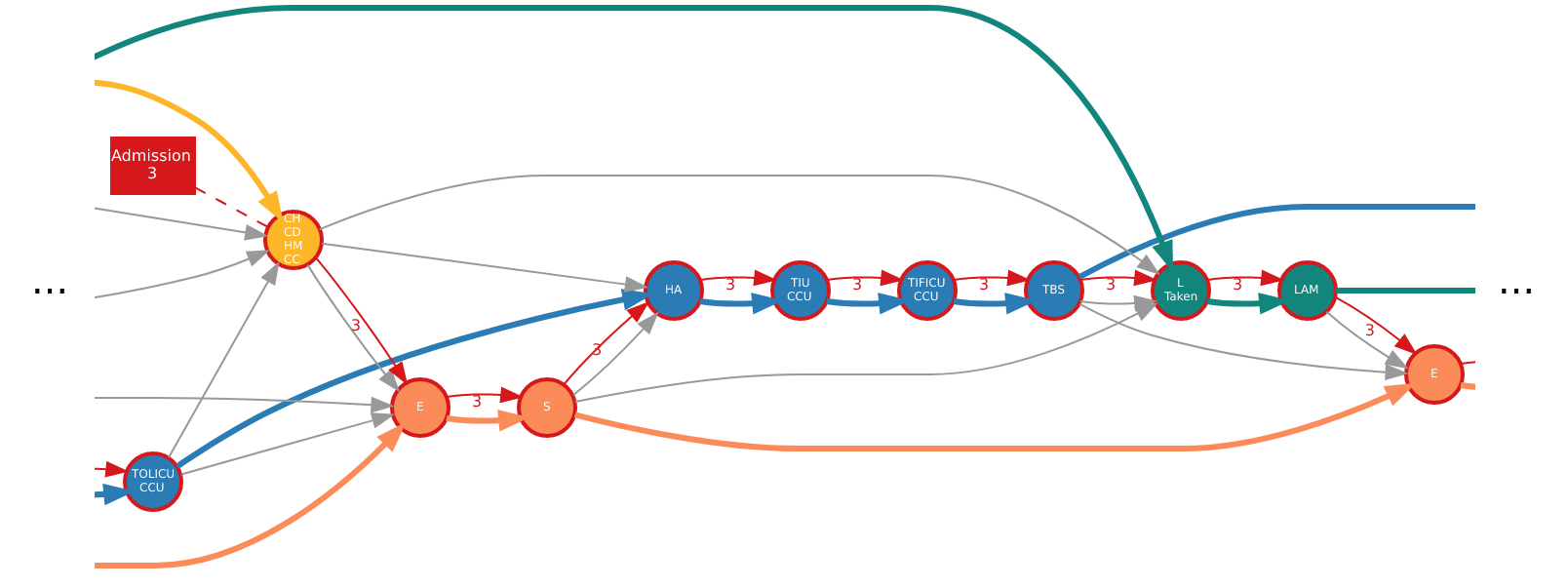}\newline
    \includegraphics[scale=0.42]{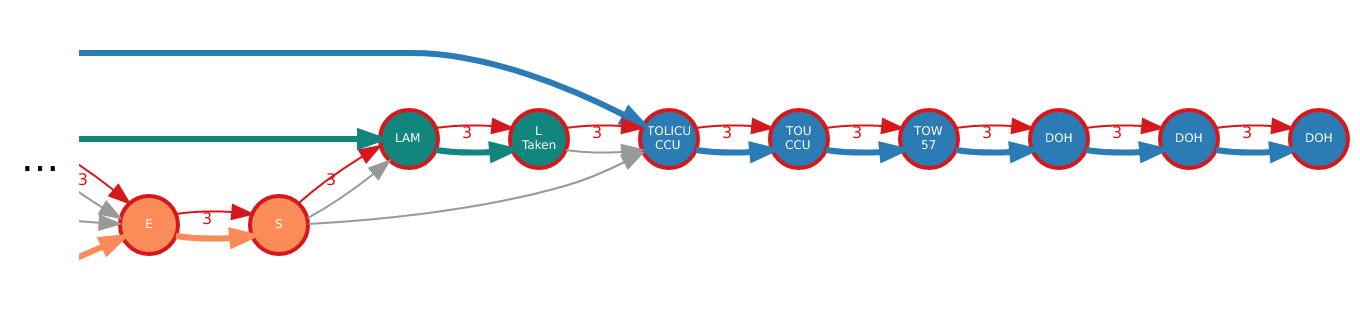}
    \caption{Multi-entity directly-follows graph for Patient\_14606 (top) and details (bottom)}
    \label{fig:two}
\end{figure}

The multi-entity directly-follows graphs of two patients are shown in Figures~\ref{fig:one} and \ref{fig:two}. These two patients, Patient\_4900 and Patient\_14606, are examples of multi-morbid patients who have been admitted to the hospital several times and had more than one disease at each time of admission. 

Based on Figure~\ref{fig:one}, before hospital admission, a laboratory measurement was taken (L Taken Node) for Patient\_4900, the abnormal measurements (LAM node) of laboratory test is one of the bases for diagnosing diseases for that patient. The patient was admitted to the hospital three times. In each of them, several diseases were diagnosed for the patient, and after that, several activities related to \textbf{Logistic}, \textbf{Laboratory\_Measurement} and \textbf{Prescriptions} entities happened for the patient. In the first, second, and third admission, respectively, four, six, and four diseases were diagnosed for the patient. The activities for \textbf{Logistic}, \textbf{Laboratory\_Measurement} and \textbf{Prescriptions} entities is different in each admission because there is difference between diagnoses diseases of each three admission. It means the activities done for patients are related to their diseases. We can see that the disease CA (Coronary Atherosclerosis) and DM (Diabetes Mellitus type II) was diagnosed in all three admissions, which indicate some common activities related to entities have occurred in all three times of admission.
On the other hand, we have diseases such as HL (Hypercholesterolemia), HH (Hemorrhage), MN (Malignant neoplasm), MF (Myocardial Infarction), which were diagnosed in only one admission time. It shows that first, there are unique activities related to entities related to this disease. Second, they were treated in hospital.

According to Figure \ref{fig:two}, the patient was admitted to the hospital without any laboratory measurement, which means that patient diagnoses related to the first admission are not related to previous measurements. For patient\_14606, a group of diseases was diagnosed in the patient's first admission: CA (Coronary Atherosclerosis), CS (Coronary Syndrome), HD (Hyperlipidemia), HM (Hypothyroidism), HT (Hypertension). After that several activities related to \textbf{Logistic}, \textbf{Laboratory\_Measurement} and \textbf{Prescriptions} entities were conducted for treating those diseases. After the first patient admission, a laboratory test was taken that was used as the basis of diagnoses for the second admission. In the second admission of Patient\_14606 another group of diseases was diagnosed: DM (Diabetes mellitus), CC (Carotid Artery Occlusion), VD (Vascular Disease), HL (Hypercholesterolemia), HM (Hypothyroidism), HT (Hypertension) since then activities related to \textbf{Logistic}, \textbf{Laboratory\_Measurement} and \textbf{Prescriptions} entities happen. In the third admission of Patient\_14606, another group of diseased were diagnosed: CH (Congestive Heart Failure), CD (Cardiac Dysrhythmia), HM (Hypothyroidism), CC (Carotid Artery Occlusion). For the Patient\_14606, we can see that diseases related to coronary disease were not diagnosed in the second and third time, indicating activities in the first admission treated these diseases. Also, diseases are repeated in all three admissions, which indicates these diseases are chronic diseases or the activities are done for the patient were not useful.

\section{Discussion}

Based on the Figures~\ref{fig:one} and \ref{fig:two}, discovered multi-entity directly follows graph for those patients show all traditional process mining concepts (e.g., sequence of activities) and for all involved clinical processes in only one graph. Meanwhile, the relationship between the different clinical processes activities that were not detectable in traditional models was clearly shown in discovered directly follows graph. This graph shows how diagnoses for multi-morbid patients evolved during the care pathways and how these diagnoses relate to other events, and how the trajectory of patients varies for each group of diagnoses.

The multi-entity directly-follows graph of \textbf{Patients\_4900} and \textbf{Patient\_14606} involves four entities which each of which has been shown with different colors. Before the first Admission of the \textbf{Patients\_4900}, the patient had abnormal values related to out-of-hospital laboratory measurements from clinics which the patient had visited. These measurements can be one of the bases for diagnosing diseases for the first Admission of that patients. These diseases were shown in \textbf{Diagnoses} entity. Meanwhile, in discovered graphs, the admission number of patients was indicated by separate red edges.

These graphs demonstrate that analyzing care pathways of patients with multi-morbidity is completely applicable using an event graph. The discovered graphs for distinct patients can illustrate all single-entity concepts such as activities, cases, and their properties for all entities simultaneously. Based on these results, the hypotheses of the research, applying event graphs produce valuable insights when using multi-entity event data for clinical pathways of multi-morbid patients, seems to be valid.

\section{Conclusions}

In this research, we could discover insightful graphs comparing traditional process mining by using multi-entity event data stored in an event graph. We evaluate the potential of the event graph approach proposed by Essser and Fahland~\cite{Fahland} for clinical data by using the MIMIC-II database.
Some of the limitations of this paper are related to the case study, such absence of resources in the MIMIC-III database and shifting times. Another limitation is related to missing visualization methods for multi-entity event data. Creating appropriate visualization approaches and automating process discovery can be future research. Enabling to show sub-processes inside an event is a highly insightful capability for graphs, which can be future work. As well, multi-entity graph notations need to be researched and created. 

%
%
%
%


\end{document}